\documentclass[twocolumn,pra]{revtex4}

\usepackage{epsf}

\begin{document}
\preprint{}
\title{Singular behavior of an entangled state 
for a one-dimensional quantum spin system }
\author{Akira Kawaguchi and  Kaoru Shimizu}
\address{NTT Basic Research Laboratories, 
NTT Corporation, 
3-1 Morinosato-Wakamiya, Atsugi, Kanagawa 243-0198
}
\date{\today}
\begin{abstract}
We studied the entangled state for a one-dimensional $S=1/2$ 
antiferromagnetic quantum spin chain in a transverse field. 
We calculate the ground state using the density matrix renormalization group 
and discuss how the entangled state changes around a quantum phase 
transition (QPT) point.
By analyzing concurrence $C(\rho)$ for two-qubit density matrix $\rho$ 
after the Lewenstein-Sanpera decomposition, 
 $\rho=\Lambda \rho_s + (1-\Lambda) \rho_e $ , 
where $\rho_s$ is a separable density matrix and 
$\rho_e$ is a pure entangled state obtained 
by a linear combination of Bell states, 
we find singular behaviors both in $C(\rho_e)$ and $1-\Lambda$ 
at the QPT point.　
$C(\rho_e)$ includes the effects of quantum fluctuations, 
which manifest the competition between the antiferromagnetic spin fluctuation 
and the effect of transverse field in the transverse Ising model. 
The quantum fluctuation shows the singular maximum at the QPT point 
as expected from the general picture of phase transition. 
In contrast, $1-\Lambda$ reveals the singular minimum at QPT point. 

\end{abstract}



\maketitle



Recently, the entanglement properties of quantum spin systems have attracted much interest \cite{Osterloh,Osborne,Vidal,Vidal2,Glaser,Mont,Roscilde1,Roscilde2} . 
One of the most interesting topics is the behavior of entanglement at  various types of quantum phase transition (QPT) that are induced by the change of an external field or coupling constant at absolute zero temperature. 
This is because we can expect that the quantum fluctuation around QPT points triggers anomalous behavior. 
To study the entanglement embedded in the quantum fluctuation, it is useful to analyze the concurrence \cite{Wootters} for two-neighboring spins of the ground state. 
In practice, Osterloh {\it et al.} have shown that the derivative of the concurrence with respect to a transverse magnetic field diverges at the transition point \cite{Osterloh}. However the concurrence itself does not show a maximum. 
At a glance, this seems to go against the general picture of phase transition because the fluctuation becomes the maximum at the critical point in general. 
The reason the maximum point shifts from the critical point is not known precisely.

To give an answer to the above problem, in this paper, we apply the Lewenstein-Sanpera decomposition \cite{Lewenstein} to analyze the pairwise-entanglement of one-dimensional (1D) quantum spin systems. 
Lewenstein and Sanpera have shown that any bipartite density matrix $\rho$ can be represented optimally as a sum of a separable state $\rho_s$ and an entangled state $\rho_e$.
For two-qubit systems, they have also shown that the decomposition reduces to a mixture of a mixed separable state and an entangled pure state ($\rho_e=| \Psi_e \rangle \langle\Psi_e |$) as $\rho=\Lambda \rho_s + (1-\Lambda)\rho_e$, thus all non-separability content of the state is concentrated in the pure entangled state.
This leads to an unambiguous measure of entanglement for any two qubit state as entanglement of the pure state $| \Psi_e \rangle \langle\Psi_e |$ multiplied by the weight $(1-\Lambda)$ of pure part in the decomposition.

We have employed the density matrix renormalization group (DMRG) method  \cite{White1} to calculate the ground state of the Hamiltonian (\ref{Hamilt}) of 1D system. 
In particular, we make use of the infinite DMRG method in order to compute 
bipartite density matrixes in the bulk limit. 
The infinite DMRG enables us to treat 1D systems with high accuracy 
even around QPT point \cite{Hieida}.
But, for the Ising case ($J_{\parallel}=0$), since the exact ground state was obtained by previous works \cite{XYZ},  we have taken advantage of the exact solution to check our numerical results. 
Our numerical results clarify that (i) the concurrence $C(\rho)$ for the pure entangled state $| \Psi_e \rangle \langle\Psi_e |$ shows a singular maximum at the QPT point and (ii) the weight $(1-\Lambda)$ of the pure part shows a singular minimum at the point. 
This is the reason that the maximum point of $C(\rho)= (1-\Lambda)C(\rho_e)$ shifts from the QPT point.

\begin{figure}[t]
\begin{center}
\vspace{-0.cm}
\hspace{-0.0cm}
\leavevmode \epsfxsize=80mm
\epsffile{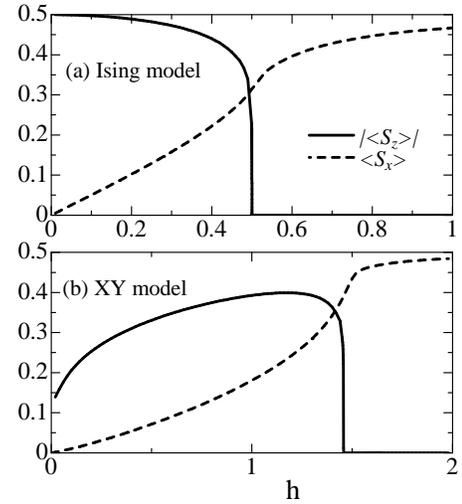}
\vspace{-5.0cm}
\end{center}
\caption{ The uniform and staggered magnetization: 
(a) Ising model ($J_{\parallel}=0,J_{\perp}=1$) and 
(b) XY model ($J_{\parallel}=1, J_{\perp}=0$).
}
\label{Mag}

\end{figure}



The Hamiltonian for the 1D $S=1/2$ anisotropic Heisenberg model in a transverse field is given by 
\begin{eqnarray}
{\cal H}&=&
      \sum_{i}[
       J_{\parallel} (S^{x}_{i}S^{x}_{i+1}
   +      S^{y}_{i}S^{y}_{i+1} )
   +   J_{\perp} S^{z}_{i}S^{z}_{i+1} ] 
\nonumber \\
&&    - h\sum_{i} S^{x}_{i}
\label{Hamilt}
\end{eqnarray}
where $S_{i}^{\alpha}$ is the $S=1/2$ spin operator at site $i$, 
and $h$ is the strength of the transverse magnetic field. 
The exchange coupling constant in the easy-plane ($xy$-plane) 
is denoted by $J_{\parallel}(\geq 0)$ and that of the easy-axis ($z$-axis) 
is $J_{\perp}(\geq 0)$.

The properties of Hamiltonian (\ref{Hamilt}) have been studied 
for several years. 
When $h=0$, the ground state of this system is described as follows. 
For $J_{\parallel} < J_{\perp}$, the systems belong to 
the Ising universality class, 
then the property of the systems are characterized with 
the spontaneous staggered magnetization $|\langle S_z \rangle|$
that has a finite value even at $h=0$. 

On the other hand, for $J_{\parallel} \geq J_{\perp}$, 
the systems belong to the Tomonaga-Luttinger liquid phase, 
therefore there is no spontaneous staggered magnetization. 
The property of this phase is that the correlation length $\xi$ of spin fluctuation 
diverges in the whole region. 
Namely, in $1D$ $S=1/2$ spin systems, the critical behavior appears in the Tomonaga-Luttinger liquid phase and in the phase transition points as well.

Then, it should be mentioned regarding, magnetic properties, there is a special value of the transverse field called a classical point $h_{cl}$. 
This point is exactly known as 
$h_{cl}=\sqrt{2J_{\parallel} (J_{\parallel}+J_{\perp}) }$, 
and the ground state can be expressed by a simple product state \cite{XYZ,Hieida}. 

In Fig. \ref{Mag}, we show the uniform and staggered magnetization 
for the Ising model and XY model. 
The classical point corresponds to $h=0$ and $h=\sqrt{2}$, respectively. 
Here, two magnetizations have the relation 
$\sqrt{ \langle S_x \rangle^2 + \langle S_z \rangle^2 }=1/2$ 
at these points.


To analyze the pairwise-entanglement of two-neighboring spins, 
we prepare the two-qubit density matrix from the wave function of 
the ground state obtained by DMRG. 
In $S=1/2$ systems, 
Wootters \cite{Wootters} has shown that the pairwise-entanglement (entanglement of formation) of a mixed state can be defined as 
\begin{eqnarray}
C(\rho)&=&\max[0, \lambda_1 -\lambda_2 -\lambda_3 -\lambda_4]
\label{Conc}
\end{eqnarray}
where the $\lambda_{\alpha}$ are the square roots 
of the eigenvalues of the matrix 
$R=\rho(\sigma_y\otimes\sigma_y)\rho^{\ast}(\sigma_y\otimes\sigma_y)$ 
in decreasing order. 
The concurrence varies from $C=0$ for a separable state 
to $C=1$ for a maximally entangled state.

Moreover, according to Lewenstein and Sanpera \cite{Lewenstein}, any two-qubit inseparable density matrix $\rho$ can be decomposed as 
\begin{eqnarray}
\rho&=&\Lambda \rho_s + (1-\Lambda)\rho_e
\label{Sept1}
\end{eqnarray}
where $\rho_s$ is a separable density matrix 
and $\rho_e=| \Psi_e \rangle \langle\Psi_e |$ 
is a pure entangled state   
obtained by a linear combination of Bell states,
$( |\psi_{\pm}\rangle =\frac{1}{\sqrt{2}} |\uparrow\downarrow\rangle
                     \pm |\downarrow\uparrow\rangle, 
|\phi_{\pm}\rangle =\frac{1}{\sqrt{2}} |\uparrow\uparrow\rangle
                     \pm |\downarrow\downarrow\rangle)$
\begin{eqnarray}
 |\Psi_e \rangle= a|\psi_{-}\rangle +b|\psi_{+}\rangle
                 +c|\phi_{-}\rangle +d|\phi_{+}\rangle
\label{Sept2}
\end{eqnarray}
where $|a|^2+|b|^2+|c|^2+|d|^2=1$. 

The Lewenstein-Sanpera (L-S) decomposition of a given density matrix $\rho$ is not unique and, in general, there is a continuum set of decompositions to choose from. The optimal decomposition is, however, unique for which $\Lambda$ is maximal. 
To determine the maximum $\Lambda$, we perform calculations as follows. 
First, we make 
\begin{eqnarray}
\tilde{\rho_s}&=&\{ \rho - (1-\Lambda)\rho_e \} /\Lambda .
\label{Sept1}
\end{eqnarray}
with variables $(\Lambda, a, b, c, d)$. 
Then we make use of the positive partial transpose (PPT) \cite{Peres,Horodecki} 
to determine whether $\tilde{\rho_s}$ is separable 
($\tilde{\rho_s}=\sum_i p_i \rho_A\otimes\rho_B$) or not. 
We find the optimum $\Lambda$, 
which is the maximum value under the condition 
that $\tilde{\rho_s}$ is separable.

If $\Lambda$ is optimum, the concurrence $C(\rho)$ for $\rho$ is related to $1-\Lambda$ and $\rho_e$ of the L-S as 
\begin{eqnarray}
C(\rho)&=& (1-\Lambda)C(\rho_e)
\label{Sept2}
\end{eqnarray}
where $C(\rho_e)=|a^2-b^2-c^2+d^2|$.


\begin{figure}[t]
\begin{center}
\vspace{-0.cm}
\hspace{-0.0cm}
\leavevmode \epsfxsize=80mm
\epsffile{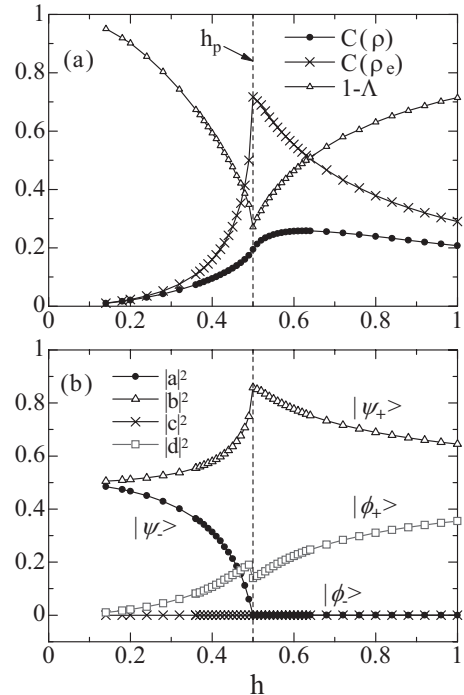}
\vspace{-1.5cm}
\end{center}
\caption{ Numerical results of L-S decomposition for 
the transverse Ising model $(J_{\parallel}=0, J_{\perp}=1)$ 
as a function of the transverse magnetic field $h$. 
$h=0.5(\equiv h_p)$ denotes the QPT point from the AFM phase to the PM phase. 
}
\label{Ising}
\end{figure}
\begin{figure}[tb]
\begin{center}
\vspace{-0.cm}
\hspace{-0.0cm}
\leavevmode \epsfxsize=80mm
\epsffile{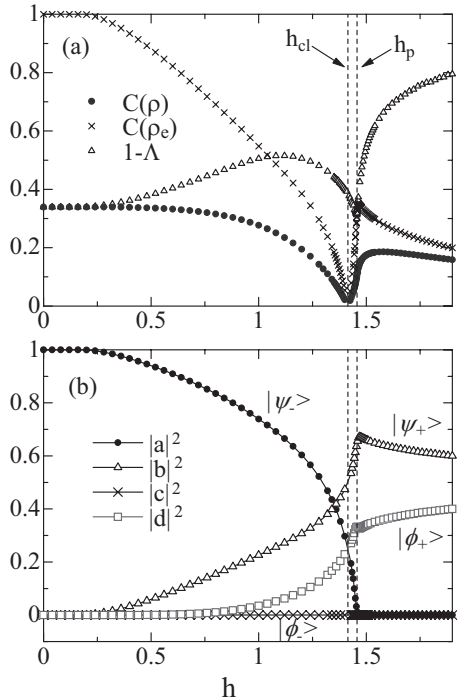}
\vspace{-1.5cm}
\end{center}
\caption{ Numerical results of L-S decomposition for 
the XY model $(J_{\parallel}=1, J_{\perp}=0)$. 
$h\simeq 1.458(\equiv h_{p})$ denote the QPT point from the AFM phase to the PM phase and $h= \sqrt{2}(\equiv h_{cl})$ is the classical point, which can be expressed by a simple product state. 
Note that the Tomonaga-Luttinger liquid state appears at $h=0$. 
}
\label{XY}
\end{figure}

In Fig. \ref{Ising}(a) and Fig. \ref{XY}(a), we show 
the values of $C(\rho)$, $C(\rho_e)$ and $1-\Lambda$ for the Ising model and XY model, respectively. 
For the Ising case, $h_p=0.5$ corresponds to the phase transition point. 
In this result, we find $C(\rho_e)$ and $1-\Lambda$ show singular behaviors at $h_p$. 
Also, for the XY model, $C(\rho_e)$ and $1-\Lambda$ has the singular maximum  and minimum at $h_p$, respectively. 
As a consequence of these contrasting behaviors, the maximum point of concurrence $C(\rho)$ shifts from the QPT point.



We consider the detail of the $C(\rho_e)$, which is the concurrence of a pure entangled state.
In Fig. \ref{Ising}(b) and Fig. \ref{XY}(b), we show the ratio of Bell states. 
The $|a|^2$ indicates the probability of the $|\psi_{-}\rangle$ state. 
This probability vanishes entirely at $h_p$ in both cases. Therefore, we can consider this quantity to be the parameter characterizing the QPT from the antiferromagnetic (AFM) phase to the paramagnetic (PM) phase. Also this is the order parameter like the staggered magnetization. 
On the other hand, $|c|^2$ means the effect of ferromagnetic fluctuation, and $|c|^2=0$ for these cases. 
But, if we consider the pairwise entanglement of next-nearest neighbor spins, 
$|c|^2$ has finite values and behaves itself like the order parameter, substituting for $|a|^2$.

In Fig. \ref{XY}(b), the point of $h=0$ is the Tomonaga-Luttinger liquid state with  zero staggered magnetization. 
In the low-field region, the system is under the influence of the AFM quantum fluctuation due to the XY-term ($J_{\parallel} \sum [S^x_i S^x_{i+1} +S^y_i S^y_{i+1}] $), while, in the high-field region, the effect of the transverse-field $h$ becomes dominant. 
In the latter case, the effect of transverse-field is reflected in $|b|^2$ and $|d|^2$. 
Here it should be noted that the total-$S^z$ is not a conservative quantity for the anisotropic models in the transverse field. Therefore, we can regard $|b|^2$ and $|d|^2$ to reflect the spin-number fluctuation. 
In particular, $|b|^2$ increases in the middle-field region. 
This means that the enhancement of $C(\rho_e)$ at $h_p$ is attributed to the spin-number fluctuation. 

Next we focus on the classical point $h_{cl}$. 
However, we can not determine the set of $(\Lambda, a, b, c, d)$ uniquely at $h_{cl}$, because a given density matrix $\rho$ is already separable. 
Therefore, we discuss the classical point from the limit of both sides of $h_{cl}$.  
As $|c|^2=0$ and $|b|^2 > |d|^2$ in all regions, 
$C(\rho_e)$ is represented by $C(\rho_e)=|a|^2-(|b|^2-|d|^2)$. 
The first term and second term imply the effect of the AFM fluctuation 
and the spin-number fluctuation, respectively. 
This equation means that the two quantum fluctuations conflict with each other. 
Actually, in Fig. \ref{XY}, $C(\rho_e)$ goes to zero at $h_{cl}= \sqrt{2}$, and then $|a|^2 \rightarrow 1/4$ , $|b|^2 \rightarrow 1/2$ and $|d|^2 \rightarrow 1/4$, respectively, 
while, for the Ising model, the classical point is $h=0$. 
Then, the entangled state is approached as  
\begin{eqnarray}
| \Psi_e \rangle \rightarrow \frac{1}{\sqrt{2}}|\psi_{-}\rangle 
                              \pm\frac{1}{\sqrt{2}}|\psi_{+}\rangle.  
\end{eqnarray}
Here the $\pm$ indicates that the ground state approaches 
$|\uparrow\downarrow\rangle$ or $|\downarrow\uparrow\rangle$, 
due to the spontaneous symmetric breaking.



We discuss the behavior of $1-\Lambda$ at around $h_p$. 
As shown in Fig. \ref{Ising}(a) and Fig. \ref{XY}(a), $1-\Lambda$ exhibits the singular minimum at $h_p$.
Moreover, as shown in Fig. \ref{XY}(a), $1-\Lambda$ shows another minimum at the point of $h=0$, where the system behaves as Tomonaga-Luttinger liquid. 
Note that these two points correspond to the critical point of the system, where the correlation length $\xi$ of spin fluctuation is infinite. 
These facts suggest that the divergence in the number $N$ of the correlated spins brings the minimum in $1-\Lambda$ observed at the critical points. 
In practice, an amount of the pairwise entanglement decreases necessarily when the number of entangled qubits increases. 



In conclusion, we have studied the pairwise-entanglement in 1D $S=1/2$ AFM quantum spin system in a transverse field. 
We have found that not only the derivative of the concurrence, which have shown by Osterloh, {\it et. al.} \cite{Osterloh}, but also the decomposed components indicate the singular behaviors at QPT point. 
We also expect these components show the scaling behaviors in the vicinity of the phase transition point. 
In particular, the probability of $|\psi_{-}\rangle$ is suitable for describing QPT from the AFM to the PM phase, because it vanishes entirely at QPT point.

We would like to thank Y. Tokura, N. Imoto, F. Morikoshi and  N. Kawakami 
for valuable discussions. 
This work was partly supported by SORST-JST. 


\end{document}